Research Article

Moisés H. Ibarra Miranda, Lars W. Osterberg, Dev H. Shah, Kartik Regulagadda and Lisa V. Poulikakos*

# 3D-architected gratings for polarization-sensitive, nature-inspired structural color

**Abstract:** Structural coloration, a color-generation mechanism often found in nature, arises from light-matter interactions such as diffraction, interference and scattering, with micro- and nanostructured elements. Herein, we systematically study anisotropic, 3D-architected grating structures with polarization-tunable optical properties, inspired by the vivid blue of *Morpho* butterfly wings. Using two-photon lithography, we fabricate multilayered gratings, varying parameters such as height (through scanning speed and laser power), periodicity, and number of layers. In transmission, significant color transitions from blue to brown were identified when varying structural parameters and incident light polarization conditions (azimuthal angle and ellipticity). Based on thin film diffraction efficiency theory in the Raman-Nath regime, optical characterization results are analytically explained, evaluating the impact of each parameter variation. Overall, these findings contribute to technological implementations of polarization-sensitive, 3D-architected gratings for structural color applications.

## 1 Introduction

Structural coloration, generated by the interaction of light with micro- and nanostructured materials, is abundant in nature [1], [2]. For instance, the *Morpho* butterfly wing incorporates three-dimensionally (3D) architected and ordered lamellas layers to produce a saturated blue [3]. Similarly, the bright green color exhibited by jeweled beetle shells arises from an arrangement of hexagonal microcells [4]. These natural photonic crystals are composed of periodically arranged optical structures that exhibit iridescent colors, which vary with the incident angle or polarization of light [5], [6]. Meanwhile, photonic glasses generate non-iridescent structural colors with a disordered or amorphous structural arrangement [7], [8]. Due to their tunability, stability, and sustainability [9], structural colors have drawn attention for applications including high-resolution image generation [10]–[12], sensing [13]–[15], displays [16]–[19], information storage [20]–[22], and anticounterfeiting [23]–[26], among others. Efforts directed toward polarization-sensitive structural color promise targeted, all-optical applications including tunable displays [27]–[29], and microstructural imaging in biological tissues and fibrous materials [30]–[32].

To enable these technological applications, artificially engineered structural colors have been demonstrated with versatile material platforms. Optical metasurfaces, composed of two-dimensionally (2D) arranged sub-wavelength elements which manipulate light-matter interactions [33], have generated structural color from resonant metallic [34]–[36], dielectric [37]–[39] or hybrid nanostructures [40]–[42]. These building blocks can be designed to exhibit optical anisotropy, e.g. *via* symmetry-breaking in their geometric design, which results in a structural color response that is tunable with the polarization state of light [43], [44].

However, complex and costly top-down fabrication techniques, including electron beam lithography, are often required to manufacture the sub-wavelength features that comprise colorimetric metasurfaces, particularly regarding the anisotropic building blocks needed for incident angle- and polarization-sensitive applications [45]–[49]. Thus, the high costs of material and equipment, as well as scalability, can pose prohibitive barriers to large-scale applications [50]. In contrast, 3D-architected photonic crystals, where structural color is achieved by larger feature sizes than 2D metasurfaces [51], promise opportunities to overcome fabrication challenges while operating at lower material cost [52]–[54]. Therein, color generation is enabled by additional degrees of freedom in light modulation, including diffraction, interference, and scattering, rather than single-particle resonances and material properties [51], [55], [56]. Indeed, many natural structural colors arise from 3D-architected refractive-index variations of chitin [57], yet artificial replicates, particularly where polarization-sensitive response is required, remain challenging [51], [58].

Two-photon lithography (TPL) (alongside greyscale photolithography [59]–[62] and colloidal self-assembly [63]–[67]) has emerged as a promising technique for additive manufacturing of 3D-architected photonic crystals [68]. Liu *et al.* [5] explored the structural coloration generated by complex 3D systems fabricated *via* commercial TPL equipment, coupled with a heat-induced shrinking process. These showed photonic stopbands over the visible range and vibrant colors dependent on the lattice constants of the structure, approaching sizes as small as 280 nm. The study indicated the potential of TPL to realize dielectric 3D-architected systems for structural color. Similarly, Gu *et al.* [69] implemented TPL to fabricate biomimetic structural colors, retrieving inspiration from lamella arrangement found in the *Morpho* butterfly. Therein, the authors demonstrated the ability to vary geometry by controlling TPL parameters such as laser power, scan speed, and interface position, to generate different colors. Simultaneously, they identified multilayer- and thin-film-interference as the main coloration mechanisms. Cao *et al.* [51] studied bi-grating structures fabricated *via* TPL, retrieving inspiration from *Cynandra opis* butterfly wings. Structural coloration was analyzed under alterations of periodicity, height, and incident light angle, which contributed to the tuning of the resulting color throughout the visible region. Fabrication parameters influenced the multilayer interference and diffraction that ultimately resulted in pure and tunable coloration. Likewise, 3D woodpile photonic crystals were fabricated by Liu *et al*. [70] with TPL, highlighting the anisotropic photonic stop band exhibited by the geometry to generate controllable vivid structural coloration. They achieved 3D profiles by controlling the scan speed and laser power during the TPL fabrication step, thus adjusting the color. While these studies showcase the realization and versatility of 3D-architected structural color systems fabricated by TPL, the potential of their structural and optical anisotropy for tunable color generation remains largely unexplored.

*\*Corresponding author: Lisa V. Poulikakos**, Department of Mechanical and Aerospace Engineering, Program of Materials Science and Engineering, University of California San Diego, La Jolla, CA, USA; E-mail: lpoulikakos@ucsd.edu; https://orcid.org/0000-0002-1118-789X
**Moisés H. Ibarra Miranda:** Program of Materials Science and Engineering, University of California San Diego, La Jolla, CA, USA; E-mail: moibarramiranda@ucsd.edu; https://orcid.org/0009-0008-7685-974X
**Lars W. Osterberg:** Department of Mechanical and Aerospace Engineering, University of California San Diego, La Jolla, CA, USA; E-mail: losterberg@ucsd.edu
**Dev H. Shah:** Department of Mechanical and Aerospace Engineering, University of California San Diego, La Jolla, CA, USA; E-mail: deshah@ucsd.edu
**Kartik Regulagadda:** Laboratory for Multiphase Thermofluidics and Surface Nanoengineering, Department of Mechanical Engineering, University of California, Berkeley, CA, USA; E-mail: regkartik@berkeley.edu; https://orcid.org/0000-0002-5787-9234

The present study investigates the optical properties of anisotropic, 3D-architected gratings and their potential to generate polarization-tunable structural color, which varies with their geometric design. We explore structural colors generated by transmission multilayer gratings, inspired by the saturated blues of the *Morpho* butterfly wing. We provide a systematic analysis of the implications of light polarization and structural parameters on the observed color generation. Multilayer grating structures were fabricated using two-photon lithography, where we varied height, determined by the scanning speed and laser power, the number of layers and the periodicity. Optical characterization demonstrated significant color changes, shifting from blue to brown, when incident light polarization and structural parameters (periodicity, number of layers and height) were varied. The impact of these variations is further supported by an analytical transmittance model of multilayer structures and grating diffraction. The findings presented here demonstrate the potential of 3D-architected gratings for versatile polarization-tunable technological applications of structural color.

## 2 Analytical model

To model the transmission behavior of 3D architected gratings with incident linearly polarized (LP) light, we must correctly classify the grating in question. Bragg diffraction occurs for a thick, volume grating that exhibits one single diffracted beam only near a specific angle of incidence, while a thin, 2D grating exhibits multiple diffracted beams in the Raman-Nath diffraction regime [71]. This classification can be determined by several parameters [72], [73], while herein we consider the parameter $\rho$ which more reliably classifies the diffraction regime of gratings under computer simulations [74]. This parameter, originally derived for acoustooptics but henceforth used for gratings, is defined as

$$\rho = \frac{\lambda^2}{\bar{n}\, \Delta n \Lambda^2} \quad (1)$$

where $\lambda$ is the wavelength of the incident light, $\bar{n}$ is the average refractive index, $\Delta n$ is the refractive index modulation, and $\Lambda$ is the periodicity [75]. $\rho < 1$ classifies a grating as thin (Raman-Nath regime) while $\rho \gg 1$ classifies as thick (Bragg regime). Despite the naming of "thick" and "thin", this parameter does not depend on the physical thickness of the grating. The Fraunhofer approximation can be used for thin, Raman-Nath gratings. Meanwhile, Kogelnik's coupled wave theory is accurate for thick gratings, but only near the Bragg angle of incidence, and for all other angles the diffraction intensity is significantly diminished [76], [77]. Nevertheless, for all gratings, the diffraction behavior can be simulated using rigorous coupled-wave analysis [78].

The unmodified gratings analyzed in this work generally exhibit $\Delta n$, $\bar{n}$, and $\Lambda$ on the order of 0.5, 1.1, and 1.1 µm, respectively. These values result in $\rho < 1$, and the Raman-Nath regime will apply for most visible wavelengths (400-700 nm). Note that as we modify the grating parameters this may no longer be the case. Indeed, when $\Lambda$ is decreased, the wavelengths of light for which the Raman-Nath classification holds ($\rho < 1$) become smaller (see Supplementary Material, Section 1 for details). For multiple diffracted beams, the Fraunhofer far field scalar approximation is defined as

$$E_m = \frac{1}{\Lambda} \int_0^\Lambda E_i \exp\left[2\pi i \left(\frac{mx}{\Lambda} + \frac{dn(x)}{\lambda}\right)\right] dx \quad (2)$$

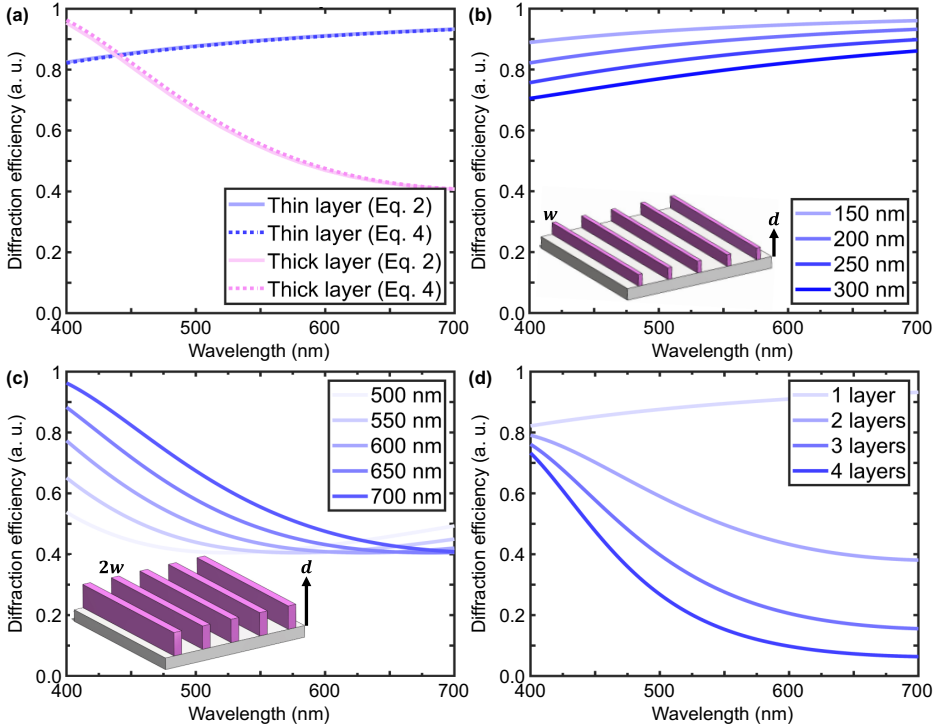

**Figure 1.** (a) Comparing the analytical diffraction efficiency results of Equation 2 with Equation 4 for a physically thin grating (pillar height $d$=200 nm) and a physically thick grating ($d$=700 nm). (b,c) Analytical diffraction efficiency for varying $d$ from (part b) 150 to 300 nm, with pillar width $w$=100 nm and grating periodicity $\Lambda$ =1.1 µm, and (part c) from 500 to 700 nm, with $2w$=200 nm and $\Lambda$=1.1 µm. (d) Analytical diffraction efficiency for stacks of 1 to 4 grating layers. The first layer has $d$=200 nm, $w$=100 nm, and $\Lambda$=1.1 µm. Layers 2 and up have $d$=700 nm, $2w$=200 nm, and $\Lambda$=1.1 µm.

where $E_m$ and $E_i$ denote the $m$'th diffracted order and incident electric field amplitudes, respectively, $d$ is the grating height, and $n(x)$ is the refractive index profile across the grating [79], [80]. Since the grating profile is rectangular, $n(x)$ is defined as

$$n(x) = 1 + (n_{polymer} - 1)\text{rect}\left(\frac{x}{w} - \frac{\Lambda}{2w}\right)$$

where $w$ is the grating width. The refractive index of the polymer material ($n_{polymer}$) is wavelength dependent and given by Cauchy's equation for refractive index defined as

$$n_{polymer}(\lambda) = A + \frac{B}{\lambda^2} + \frac{C}{\lambda^4} \quad (3)$$

where $A$=1.52266, $B$=0.00733 µm², and $C$=-0.482·10⁻⁴ µm⁴ for a TPL IP-Dip photopolymer [81]. After dividing Equation 2 by $E_i$, we can perform the integration to obtain $E_m/E_i$. Then, finding the squared modulus of this result, the diffraction efficiency ($\eta_m = |E_m|^2/|E_i|^2$) for the $m$'th diffracted order can be found for any wavelength.

The efficiencies calculated via (2) can also be expressed by

$$\eta_0 = 1 - \frac{2w}{\Lambda} + \frac{2w^2}{\Lambda^2} + \frac{2w}{\Lambda}\left(1 - \frac{w}{\Lambda}\right)\cos\left(\frac{2\pi}{\lambda d \Delta n}\right) \quad (4)$$

$$\eta_m = \frac{4}{\pi^2 m^2}\sin^2\left(\frac{\pi m w}{\Lambda}\right)\sin^2\left(\frac{\pi}{\lambda d \Delta n}\right) \quad (5)$$

[82]. For simplicity, we only consider zeroth order diffraction efficiencies in (4) due to the limited angles of admission by the numerical aperture of the objective used for imaging and experimental results. Figure 1(a) compares the results of the integral method from (2) with the analytical expression of (4). The results from both methods are almost identical, confirming the validity of (4).

Next, we incorporate the modification of certain parameters such as $d$, $\Lambda$, and number of layers into the model. Their specific values, as well as that of $w$, were chosen for their proximity to the parameters describing the experimentally fabricated gratings studied in this work. Figure 1(b)-(c) shows $d$ variations in the grating (i.e. changing values of $d$ in (4)).

We see that for lower $d$ values (physically thin gratings) in Figure 1b, the diffraction efficiency is near unity and as $d$ increases, the efficiency lowers uniformly. For higher $d$ values (physically thick gratings) in Figure 1c, diffraction is more selective, and the efficiency of blue light strengthens as $d$ increases. Modifications of $\Lambda$ are shown in Figure S1, where the Raman-Nath classification with $\rho < 1$ only holds for certain $\Lambda$ values considered.

Multilayer gratings studied here, where the grating direction is orthogonal for each subsequent layer, allow us to distinguish each layer as a separate grating. In contrast, grating layers oriented parallel would theoretically combine to form a single, thicker grating. Therefore, we multiply the diffraction efficiencies of each subsequent layer to obtain the diffraction efficiency of the whole multilayer, as $\eta$ is defined as the ratio of transmitted power to incident power [76], [80]. Since we are only considering zeroth order diffraction, this simplifies our multilayer diffraction efficiency, and we can write the efficiency for the whole stack of j layers, where $\eta_0^k$ is the efficiency of the $k'$th layer, as $\eta_0^{j\ layers} = \prod_{k=1}^{j} \eta_0^k$, as shown in Figure 1(d). Despite variations in the number of layers, a strong $\eta_0^k$ of blue light is exhibited. For the thin 1 layer ($d$=200 nm), $\eta_0^1$ is uniform and peaks slightly for red light. Adding a thicker second layer ($d$=700 nm) has the effect of diminishing this efficiency at red wavelengths quite significantly in the total $\eta_0^2$. Adding layers 3 and 4 further diminishes the efficiency at red wavelengths. However, as the 1st and negative 1st diffracted orders are not incorporated, the model for 3 and 4 layers is insufficient to predict accurate diffraction efficiencies of light, as adding layers has the effect of increasing the number of diffraction orders. Nonetheless, the presented analytical model provides valuable insights on the effect of geometric parameters on $\eta$ of transmission gratings.

threshold of laser power which is only achieved at the focal volume of the illuminating femtosecond laser beam. The resulting reduced voxel size enables high-resolution, layered printing of 3D structures [84]. TPL involves a nonlinear dependency of the polymerization rate on the laser intensity [68] in contrast to conventional single-photon lithography (see Figure 2(a)). In the latter, low-intensity light, typically at ultraviolet wavelengths, interacts with the photoresist and promotes photopolymerization along the complete beam path, which significantly limits the attainable spatial resolution and capability to produce 3D geometries.

## 3 Fabrication technique

In this study, TPL was employed to fabricate grating structures with various $\Lambda$, $d$, and numbers of layers. This 3D fabrication technique has drawn attention due to its high spatial resolution and compatibility with complex micro- and nanostructures [83]. TPL leverages a nonlinear two-photon absorption process, where polymerization of the photoresist occurs above a

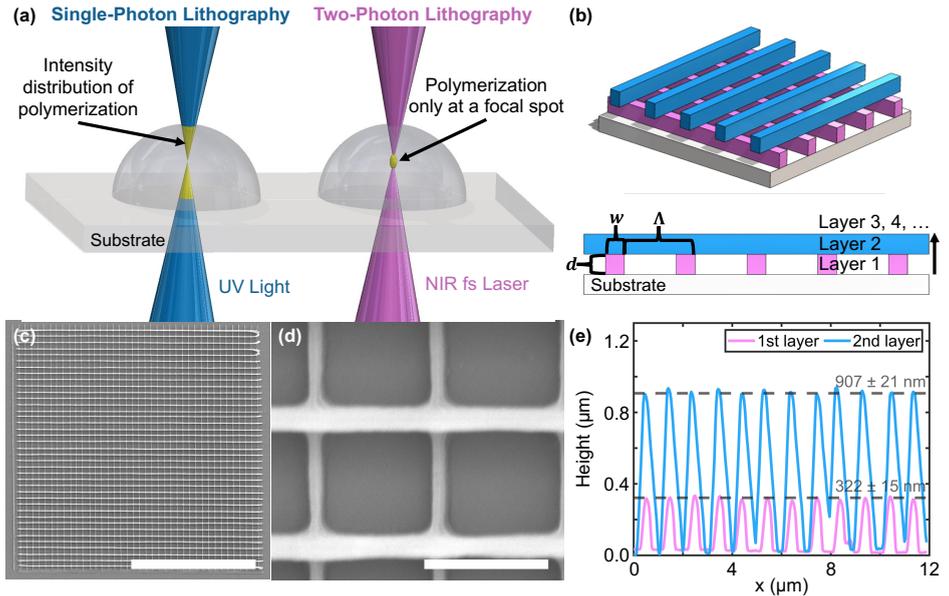

**Figure 2.** (a) Schematic of single-photon lithography (SPL, left) and two-photon lithography (TPL). (b) Design of the 3D architected bi-grating structure, characterized by pillar height and width $d$, $w$, grating periodicity $\Lambda$, and number of layers. (c,d) SEM images of a bi-grating sample with $\Lambda$=1.1 µm, top layer grating width ($w_{top}$) of 198 ± 10 nm, bottom layer grating width ($w_{bottom}$) of 122 ± 4 nm, and $d$=907 ± 21 nm; part c scale bar: 20 µm, part d scale bar: 1 µm. (e) AFM analysis of the layer heights of a bi-grating structure, showing the first layer $d$ (pink) and the overall height (blue).

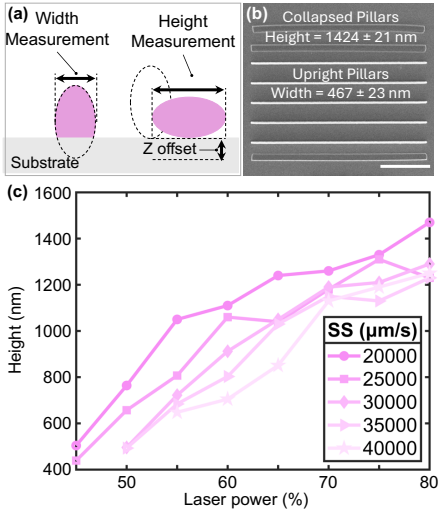

**Figure 3.** (a) Schematic of the influence of the z-offset on $d$ and $w$ printing during TPL for a standing grating (left) and a collapsed grating (right). (b) SEM image of collapsed and standing grating pillars ($d$=1424 ± 21 nm and $w$=467 ± 23 nm), using fabrication parameters of SS = 5000 µm/s and LaP = 60%. Scale bar: 10 µm. (c) Analysis of the influence of SS and LaP on pillar height $d$ as measured from collapsed gratings.

Herein, TPL was performed using *Photonics Professional GT* (Nanoscribe GmbH) and a two-photon polymerization IP-Dip photoresist was deposited on a fused silica substrate (see Supplementary Material, Section 2 for details). Figure 2(b) shows a schematic of the designed bi-grating, comprising two mutually perpendicular grating layers. Multiple grating geometries were fabricated with Λ ranging from 0.8 to 1.1 µm in 0.1-µm increments, and the number of grating layers ranging from 1 to 4. Moreover, laser power (LaP) and scanning speed (SS) were varied from 55% to 65% (with respect to a total power of 40 mW) and from 30000 to 50000 µm/s, respectively.

The morphology of the studied grating structures was characterized using an SEM *Sigma 500* (ZEISS). Figure 2(c)-(d) (scale bar of 20 µm (part c) and 1 µm (part d), respectively) shows a representative image of a bi-grating structure in a 40 µm × 40 µm area at varying magnifications, exhibiting uniformity in terms of grating spacing and direction, as well as consistent printed linewidths. These conditions were comparably found in samples with different structural parameters (see Supplementary Material, Section 3 for details). AFM studies were performed with a *Park NX20* (Park Systems) to characterize $d$. Figure 2(e) shows a $d$ profile analysis of a bi-grating structure, displaying $d$ values of 907 ± 21 nm and 322 ± 15 nm for the first and second grating layers, respectively. Additional grating samples were correspondingly characterized to study the impact of $d$ on the transmission structural coloration (see Supplementary Material, Section 4 for details). The characterization results validated the correlation between $d$ and the TPL LaP-SS input conditions.

Morphological characterization demonstrated different $w$ and $d$ values for the bottom and top grating layers. This is attributed to the generation of a z-offset during the TPL process, which is related to the calibration of the laser focus onto the photoresist layer. As shown in Figure 3(a), this offset creates a difference in the linewidth and $d$ features, which predominantly intervenes in the bottom layer. This calibration is addressed in Figure 3(b) by the fabrication and identification of non-attached collapsed pillars, which appear when the TPL voxel is generated at a location sufficiently far from the substrate-photoresist interface. This experiment yielded a $d$ value of 1424 ± 21 nm, measured for the collapsed pillars, and a $w$ value of 467 ± 23 nm, measured for the upright pillars (SS = 5000 µm/s and LaP = 60%). A similar procedure is reported by Wang *et al.* [85], which determines the proportions of $d$ and $w$ of the grating line. Additionally, $d$ was estimated by coupling specific optimized LaP-SS values (see Figure 3(c)) based on the lumped 3D printing TPL parametric model [85]. In this analysis, it is demonstrated that slower scanning speeds and increased laser powers result in larger voxel sizes and subsequent $d$. Moreover, $d$ measurements obtained from AFM and those expected from the parametric model suggest an overlap between the two layers, which is also attributed to the z-offset and stage calibration during the fabrication process. In this regard, Liu *et al.* [70] reported this overlapped structure fabricated with TPL, where woodpile photonic crystals showed superimposed stack of rods with elliptical cross sections.

Overall, TPL generated consistent grating samples for a systematic optical characterization and a reliable analysis of the impact of fabrication parameters and light polarization on transmitted structural color.

## 4 Optical characterization

### 4.1 Light polarization variation

Variation of polarization and incident light conditions provide all-optical tools to dynamically tune structural coloration [28]. Herein, the influence of light polarization conditions on structural coloration of the studied bilayer grating system is examined. Polarization control was achieved by variation of the azimuthal angle ($\varphi$) of the incident linearly polarized (LP) plane wave with respect to the grating, and by tuning of its ellipticity from LP to circularly polarized (CP) light [86]. Polarized light microscopy (Nikon ECLIPSE LV100ND, with a Nikon LV-HL 50W 12V LONGLIFE halogen lamp illumination source) was implemented to capture the transmitted color generated at a given parameter value, which was correspondingly plotted in the CIE 1931 2-degree Standard Observer color space. Simultaneously, transmittance spectra were acquired (Princeton Instruments IsoPlane 160) in the visible wavelength regime (400-700 nm) (see Supplementary Material, Section 5 for details).

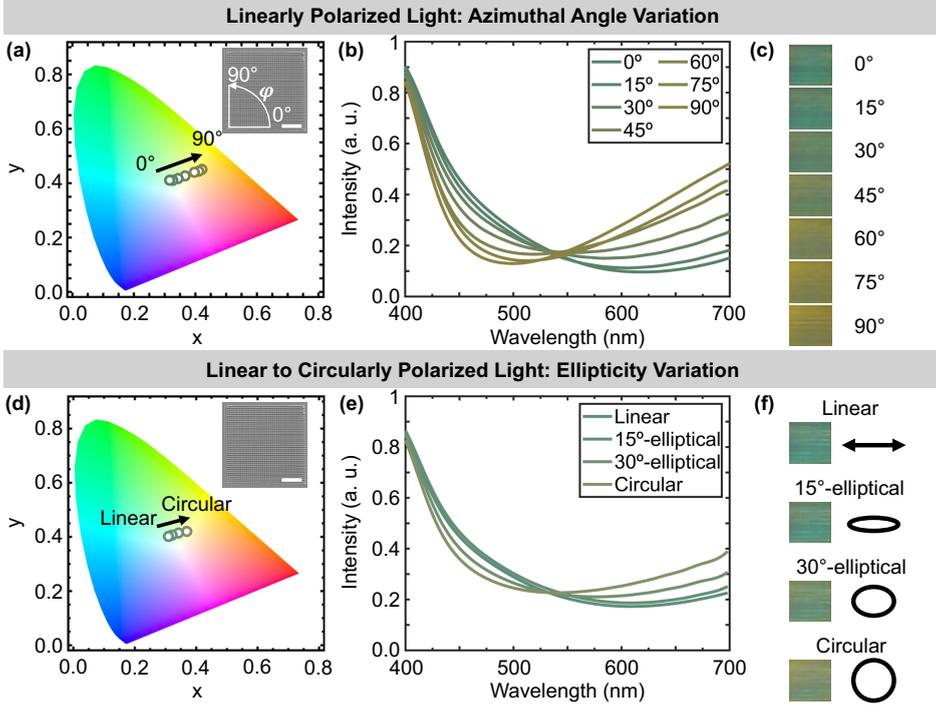

**Figure 4.** Polarization-dependent colorimetric response of a bi-grating with $\Lambda$=1.1 µm, $w_{top}$=198 ± 10 nm and $w_{bottom}$=122 ± 4 nm and $d$=907 ± 21 nm. (a)-(c) Linearly polarized (LP) light excitation with variation of the azimuthal angle ($\varphi$) from 0° to 90°. (d)-(f) LP at $\varphi$=0° to circularly polarized light excitation with variation of ellipticity. (a,d) Chromaticity diagram for transmission color response of bigrating in the CIE 1931 2-degree Standard Observer color space. Inset: SEM of the bigrating, scale bar: 10 µm. (b,e) Transmittance spectra and (c,f) optical microscope images of bigrating color response.

Herein, the structure to be analyzed is a bi-grating with $\Lambda$=1.1 µm, $d$=907 ± 21 nm, and $w$ values for the top and bottom layer of $w_{top}$=198 ± 10 nm and $w_{bottom}$=122 ± 4 nm, respectively. The first scenario, shown in Figure 4(a)-(c), involves LP incident light where $\varphi$ varies from 0º to 90º with a 15º-step. $\varphi$ = 0° corresponds to an incident LP plane wave perpendicular to the grating line direction in the bottom layer and parallel to that of the top layer. The CIE chromaticity diagram (see Figure 4(a)) shows a linear color transition, while the spectra (see Figure 4(b)) indicate a stronger transmittance at longer wavelengths with increasing $\varphi$. This variation yielded a gradual color change from blue (≤30º) to brown (see optical microscope images in Figure 4(c)). Overall, a layer becomes responsive in terms of the grating efficiency $\eta$ at a given $\varphi$. At $\varphi$=0°, we observe a dominant blue color produced by the top layer (low $\eta$ at longer wavelengths), while at $\varphi$=90° a brown color is captured due to the bottom layer diffraction (higher $\eta$ at longer wavelengths). The observed color contribution from each layer is in accordance with the analytical modeling of various $d$ (see Figure 1 (b)-(c)).

Other studies [87], [88] have focused on Bragg gratings with polarization-sensitive properties, observing a similar responsive behavior at non-zero incident angles. Therein, color responses are observed when the LP plane wave is perpendicular to the grating line direction. Interestingly, our bottom and top layers become responsive when their grating line directions are parallel to the LP light plane wave, even at normal incidence (see Supplementary Material, Section S6). Nonetheless, it is important to note that our structures are delimited within the Raman-Nath regime for thin film diffraction, defined by the parameter $\rho$ (see Equation (1)). Thus, the interpretation of the diffraction shown by these gratings will differ compared to that of gratings in the Bragg regime.

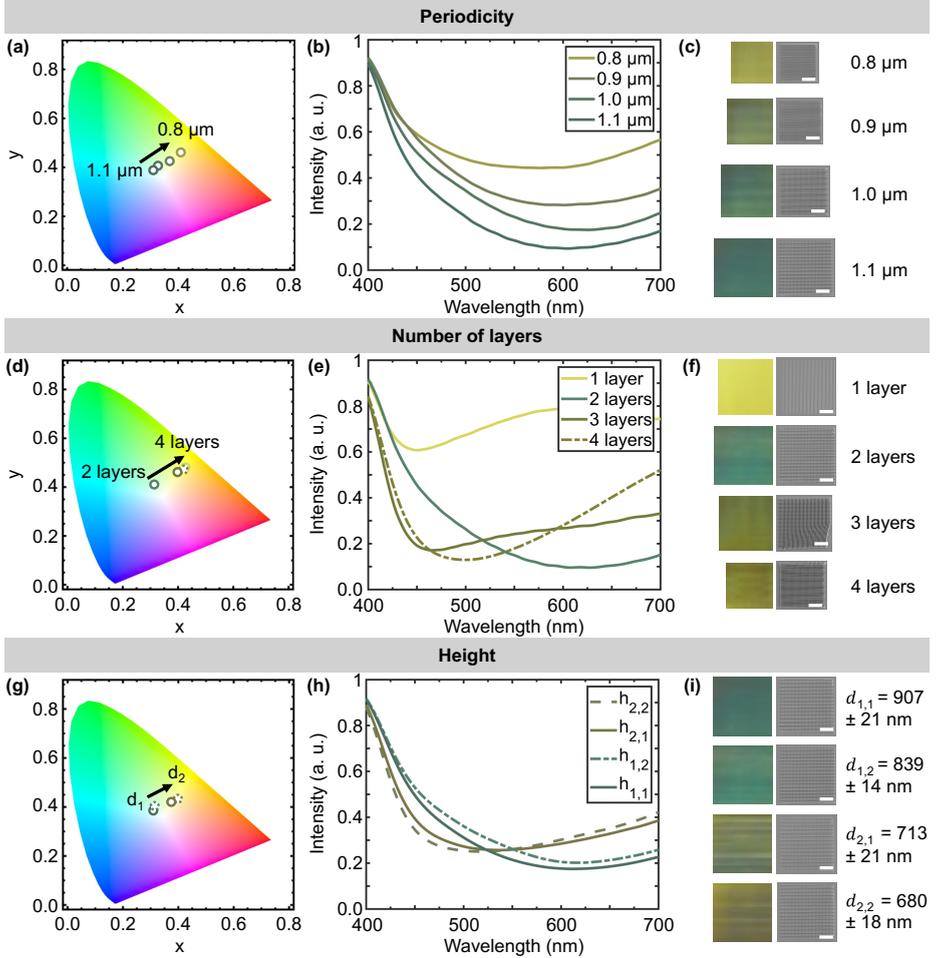

**Figure 5.** Fabrication-parameter-dependent colorimetric response of gratings for LP light excitation at $\varphi$=0°. Variations in (a)-(c) bi-grating periodicity ($\Lambda$ ranges from 0.8 µm to 1.1 µm), (d)-(f) number of layers (ranging from 1 to 4 grating layers with $\Lambda$=1.1 µm), and (g)-(i) bi-grating pillar height ($d$ ranges from 680 ± 18 nm to 907 ± 21 nm with $\Lambda$=1.1 µm) are studied. (a,d,g) Chromaticity diagram for transmission color response of gratings in the CIE 1931 2-degree Standard Observer color space. (b,e,h) Transmittance spectra and (c,f,i) optical microscope and SEM images (parts c,f) of grating color response. Scale bar: 10 µm.

Next, in Figure 4 (d)-(f), the polarization of the incident light was varied from LP to CP, with two intermediate elliptically polarized light conditions with opening angles of 15º and 30º, respectively. With increasing ellipticity, a less vibrant blue color is achieved (see CIE chromaticity diagram in Figure 4(d)) due to an increasing transmittance over longer wavelengths (see spectra in Figure 4(e)). In Figure 4(f), optical microscope images show this blue fade when transitioning from LP to CP excitation. This is attributed to the orientation of the incident light polarization, which provides equal contributions along the x- and y-axis for CP [89], similar to that observed for 45° LP excitation in Figure 4(c). Note that varying the handedness of the incident CP light would not alter the optical response due to the symmetry of the bi-grating.

### 4.2 Fabrication parameters variation

To further assess the structural color generation of the studied 3D-architected transmission grating systems, we vary structural parameters, namely $\Lambda$, $d$, and number of layers. Figure 5(a)-(c) shows the color generation of bi-gratings with $\Lambda$ values of 0.8, 0.9, 1.0, and 1.1 µm under LP light illumination at $\varphi$=0° orientation. The CIE 1931 chromaticity diagram shows a clear transition from blue to brown structural color as $\Lambda$ is decreased (see Figure 5(a)). Here, $\Lambda$=1.1 µm presents a more selective transmittance at shorter wavelengths, while a reduced $\Lambda$ exhibits increased transmittance across the visible spectrum (see Figure 5(b)), which ultimately exhibits a brown color at $\Lambda$=0.8 µm (see Figure 5(c)). Based on the aforementioned thin-film theory and the Fraunhofer approximation, $\eta$ involves a sinusoidal dependence on $\Lambda$ (see Equations (4)-(5)); thus, color response is altered as $\Lambda$ changes.

Next, Figure 5 (d)-(f) examines transmitted colors for gratings with 1, 2, 3, and 4 mutually orthogonal layers for $\varphi$=0° LP light excitation. A saturated blue color is displayed by the bi-grating structure, which eventually transitions to a brown color when incrementing the number of layers to 3 and 4 (see Figure 5(d)). This change is attributed to multilayer grating interference and additional reflections at interfaces [90], showing intense transmittance not only for blue wavelengths but also for red ones (see Figure 5(e)). In contrast, Figure 5(f) shows that a one-layer grating displays a yellow color, dominated by the halogen light source, rather than structural coloration observed for the 2-, 3- and 4-layer gratings. This correlates with the forementioned $d$ variation study in our analytical model (see Figure 1(b)-(d)) and fabrication analysis (see Figures 2(e), S2 and S4). The studied systems exhibit a thinner first grating layer ($d$ regime shown in Figure 1(b)) compared to the subsequent layers ($d$ regime shown in Figure 1(c)). Moreover, according to the thin film diffraction efficiency theory, a less wavelength-sensitive $\eta$ results as $d$ decreases. Therefore, an unsaturated color is shown for the 1-layer grating, acting as a zeroth-order transmission grating, while color saturation increases for the 2-, 3- and 4-layer gratings (see Figure 1(d)).

Figure 5(g)-(i) conducts a $d$ variation study of bi-grating structures with Λ=1.1 μm for $\varphi$=0° LP light illumination. Two pairs of bi-grating structures with the same fabrication conditions were analyzed: 1) SS of 30000 μm/s and LaP of 60% and 2) SS of 50000 μm/s and LaP of 65%. While $d$ measurements differed between pairs (see Figure S4), the optical characterization demonstrated a strong similarity in structural coloration (see Figure 5(g)), which provides information on the reproducibility of the studied TPL fabrication. Specifically, a consistent trend in the transmittance spectra was observed for different $d$, where intensities at longer wavelengths increase, as $d$ is decreased (see Figure 5(h)). Blue colors are shown in samples with $d_{1,1}$ = 839 ± 14 nm and $d_{1,2}$ = 907 ± 21 nm. Meanwhile, those bi-gratings with $d_{2,1}$ = 680 ± 18 nm and $d_{2,2}$ = 713 ± 21 nm generated brown structural color (see Figure 5(i)). This spectral behavior is in good agreement with Equations (4)-(5), where a sinusoidal relationship between $d$ and $\eta$ is stipulated. Furthermore, the observed $d$-dependent color variation follows the proposed analytical model, which indicates a stronger $\eta$ at shorter wavelengths as $d$ increases (see Figures 1(b)-(c)).

## 5 Conclusions

In this study, 3D-architected gratings with anisotropic optical responses are fabricated and systematically characterized. Based on thin film diffraction efficiency theory, optical responses by transmission gratings in the Raman-Nath regime were analytically modeled, revealing selective diffraction efficiency at blue wavelengths for appropriately designed structural parameters. Λ, $d$, and number of layers were determined by TPL fabrication conditions, such as scan speed (SS), laser power (LaP), and interface position. Optical characterization demonstrated significant color shifts from blue to brown when decreasing Λ and $d$ and increasing the number of layers. The transmittance increment at longer wavelengths was similarly observed when increasing the ellipticity from LP to CP incident light and altering the LP azimuthal angle $\varphi$. Specifically, an activation behavior where the grating effectively transmits polarized incident light at selected wavelength ranges was observed. Interestingly, the grating became responsive when the LP plane wave orientation was parallel to the grating line direction. This suggests a distinctive polarization-sensitive diffraction behavior for gratings in the Raman-Nath regime, which can be further explored in future work.

Inspired by structural colors found in nature, e.g. in blue Morpho butterfly wings, this work indicates interesting tools to systematically tune transmitted structural coloration by varying incident light polarization conditions and structural design. These results suggest the potential of anisotropic, 3D-architected gratings as polarization-sensitive structural colors in technological implementations.

**Acknowledgments:** The authors thank Saaj Chattopadhyay, Paula Kirya, Zaid Haddadin and Loren Phillips for helpful discussions.

**Research funding:** The authors gratefully acknowledge funding from the Arnold and Mabel Beckman Foundation (Beckman Young Investigator Award, Project Number: 30155266).

**Author contribution:** All authors have accepted responsibility for the entire content of this manuscript and approved its submission.

**Conflict of interest**: Authors state no conflict of interest.

**Data availability statement**: Data generated and analyzed in this study are available from the corresponding author upon request.